# Automated Website Fingerprinting through Deep Learning

Vera Rimmer[*], Davy Preuveneers[*], Marc Juarez[§], Tom Van Goethem[*] and Wouter Joosen[*]
[*]imec-DistriNet, KU Leuven
Email: {firstname.lastname}@cs.kuleuven.be
[§]imec-COSIC, ESAT, KU Leuven
Email: marc.juarez@esat.kuleuven.be

*Abstract*—Several studies have shown that the network traffic that is generated by a visit to a website over Tor reveals information specific to the website through the timing and sizes of network packets. By capturing traffic traces between users and their Tor entry guard, a network eavesdropper can leverage this meta-data to reveal which website Tor users are visiting. The success of such attacks heavily depends on the particular set of traffic features that are used to construct the fingerprint. Typically, these features are manually engineered and, as such, any change introduced to the Tor network can render these carefully constructed features ineffective. In this paper, we show that an adversary can automate the feature engineering process, and thus automatically deanonymize Tor traffic by applying our novel method based on deep learning. We collect a dataset comprised of more than three million network traces, which is the largest dataset of web traffic ever used for website fingerprinting, and find that the performance achieved by our deep learning approaches is comparable to known methods which include various research efforts spanning over multiple years. The obtained success rate exceeds 96% for a closed world of 100 websites and 94% for our biggest closed world of 900 classes. In our open world evaluation, the most performant deep learning model is 2% more accurate than the state-of-the-art attack. Furthermore, we show that the implicit features automatically learned by our approach are far more resilient to dynamic changes of web content over time. We conclude that the ability to automatically construct the most relevant traffic features and perform accurate traffic recognition makes our deep learning based approach an efficient, flexible and robust technique for website fingerprinting.

## I. INTRODUCTION

The Onion Router (Tor) is a communication tool that provides anonymity to Internet users. It is an actively developed and well-secured system that ensures the privacy of its users' browsing activities. For this purpose, Tor encrypts the contents and routing information of communications, and relays the encrypted traffic through a randomly assigned route of nodes such that only a single node knows its immediate peers, but never the origin and destination of a communication at the same time. Tor's architecture thus prevents ISPs and local network observers from identifying the websites users visit.

As a result of previous research on Tor privacy, a serious side-channel of Tor network traffic was revealed that allowed a local adversary to infer which websites were visited by a particular user [14]. The identifying information leaks from the communication's meta-data, more precisely, from the directions and sizes of encrypted network packets. As this side-channel information is often unique for a specific website, it can be leveraged to form a unique fingerprint, thus allowing network eavesdroppers to reveal which website was visited based on the traffic that it generated.

The feasibility of Website Fingerprinting (WF) attacks on Tor was assessed in a series of studies [25], [31], [19], [24], [32]. In the related works, the attack is treated as a classification problem. This problem is solved by, first, manually engineering features of traffic traces and then classifying these features with state-of-practice machine learning algorithms. Proposed approaches have been shown to achieve a classification accuracy of 91-96% correctly recognized websites [30], [24], [13] in a set of 100 websites with 100 traces per website. Their works show that finding distinctive features is essential for accurate recognition of websites. Moreover, this tasks can be costly for the adversary as he has to keep up with changes introduced in the network protocol [4], [20], [9]. The WF research community thus far has not investigated the success of an attacker who automates the feature extraction step for classification. This is the key problem that we address in this work.

An essential step of traditional machine learning is feature engineering. Feature engineering is a manual process, based on intuition and expert knowledge, to find a representation of raw data that conveys characteristics that are most relevant to the learning problem. Feature engineering proved to be even more important than the choice of the specific machine learning algorithm in many applications, including WF [12], [19].

When developing a new WF attack, prior work on WF typically focuses on feature engineering to compose and select the most salient features for website identification. Moreover, these attacks are actually defined by a fixed set of features derived from this process. Thus, these attacks are sensitive to changes in the traffic that would distort those features. In particular, deploying countermeasures in the Tor network that conceal the features is sufficient to defend against such attacks. This enables an arms-race between attacks and defenses: new attacks defeat defenses because they exploit features that had not been considered before and, conversely, new defenses are designed to conceal the features that those attacks exploited.

In this paper, we propose a novel WF attack based on deep learning. Our attack incorporates automatic feature learning and, thus, it is not defined by a particular feature set. This may be a game-changer in the arms-race between WF attacks and defenses, because the deep learning based attack is designed to be adaptive to any perturbations in the features introduced by defenses. The attack we present in this work is the first automated WF attack and it is at least as effective as the state-of-the-art, manual approaches.

The key contributions of our work are as follows:

- Our study provides the first systematic exploration of state-of-the-art deep learning (DL) algorithms applied to WF, namely feedforward, convolutional and recurrent deep neural networks. We design, tune and evaluate three models – Stacked Denoising Autoencoder (SDAE), Convolutional Neural Network (CNN) and Long Short-Term Memory (LSTM). Our DL models are capable of *automatically* learning traffic features for website recognition at the expense of using more data. Moreover, we automate the model selection to find the best network hyperparameters. We demonstrate that our DL-based WF attack reaches a high success rate, comparable to the state-of-the-art techniques.
- We reevaluate prior work on our dataset and reproduce their results. We find that state-of-the-art WF approaches benefit from using more training data, similar to DL. As a result of a systematic comparison of our novel DL-based methods to previous WF approaches for the closed and open world settings, we demonstrate comparable recognition results with slight improvements of up to 2%. Furthermore, we show that our DL attack reveals more general and stable website features than the state-of-the-art methods, which makes them more robust to concept drift caused by highly dynamic web content.
- The dataset collected for the evaluation is the largest WF dataset ever gathered to date. Our closed-world dataset consists of 900 websites, with traffic traces generated by 2,500 visits each. Our open-world dataset is based on 400,000 unknown websites and 200 monitored websites. We made the generated dataset publicly available, allowing researchers to replicate our results and systematically evaluate new (DL) approaches to WF[1].

The paper is structured as follows. In Section II, we discuss related work on WF and the use of DL. Section III presents the threat model and the capabilities an adversary has for WF. The data collection process is outlined in detail in Section IV. Section V provides a reevaluation of state-of-the-art attacks on our dataset and the overall deep learning approach and evaluation. We discuss the results and limitations of our work, as well as opportunities for future research, in Section VI. Section VII concludes by summarizing our main findings.

## II. BACKGROUND

This section reviews recent related work on Tor WF attacks relying on traditional machine learning algorithms, and the application of deep learning.

Anonymous communications systems such as Tor [11] provide confidentiality of communications and conceal the destination server's address from network eavesdroppers. However, in the last decade, several studies have shown that, under certain conditions, an attacker can identify the destination website only from encrypted and anonymized traffic.

In WF, the adversary collects traffic from his own visits to a set of websites that he is interested in monitoring, visiting each site multiple times. Next, the adversary builds a website *template* or *fingerprint* from the traffic traces collected for that site. The fingerprints are built using a supervised learning method that takes the traffic traces labeled as their corresponding site, extracts a number of features that identify the site and outputs a statistical model that can be used for classification of new, unseen traffic traces. Finally, the attacker applies the classifier on unlabeled traffic traces collected from communications initiated by the victim and makes a guess based on the output of the classifier. To be able to deploy the attack, the adversary must be able to observe the traffic generated by the victim and be able to identify the user (see Section III for more details on the threat model).

The first WF studies evaluated the effectiveness of the attack against HTTPS [8], encrypted web proxies [27], [16], OpenSSH [22] and VPNs [14] and it was not until 2009 that the first evaluation of a WF attack was performed in Tor [14]. This first attack in Tor was based on a Naive Bayes classifier and the features were the frequency distributions of packet lengths [14]. Even though their evaluation showed the attack achieved an average accuracy of only 3%, the attack was improved by Panchenko et al. using a Support Vector Machine (SVM) [25]. In addition, Panchenko et al. added new features that were exploiting the distinctive *burstiness* of traffic and increased the accuracy of the attack to more than 50%.

These works were succeeded by a series of studies that claimed to boost the attacks and presented attacks with more than 90% success rates. First, Cai et al. [5] used an SVM with a custom kernel based on an edit-distance and achieved more than 86% accuracy for 100 sites. The edit distance allowed for delete and transpose operations, that are supposed to capture drop and retransmission of packets respectively. Following a similar approach, Wang and Goldberg [31] experimented with several custom edit distances and improved Cai et al.'s attack to 91% accuracy for the same dataset.

However, these evaluations have been criticized for making unrealistic assumptions on the experimental settings that give an unfair advantage to the adversary compared to real attack settings [19]. For instance, they evaluated the attacks on small datasets and considered adversaries who can perfectly parse the traffic generated by a web-page visit from all the traffic that blends into the Tor network. Furthermore, they assume users browse pages sequentially on one single browser tab and never interrupt an ongoing page-load. Recent research has developed new techniques to overcome some of these assumptions, suggesting that the attacks may be more practical than previously expected [32].

---

[1] The dataset and implementation can be found on the following URL: https://distrinet.cs.kuleuven.be/software/tor-wf-dl/.

The three most recent attacks in the literature outperform all the attacks described above and, for this reason, we have selected them to compare with our DL-based attack. Each attack uses a different classification model and feature sets and work as follows:

**Wang-kNN [30]:** this attack is based on a k-Nearest Neighbors (k-NN) classifier with more than 3,000 traffic features. This large amount of features is obtained by varying the parameters of set of fewer feature *families*. For instance, the number of outgoing packets in spans of $X$ packets and the lengths of the $Y$ packets in the same direction. In order to mitigate the curse of dimensionality, they proposed to weigh the features of a custom distance metric, minimizing the distance among traffic samples that belong to the same site. Their results show that this attack achieves 90% to 95% accuracy on 100 websites [30].

**CUMUL [24]:** CUMUL is based on an SVM with a Radial Basis Function (RBF) kernel. CUMUL uses the cumulative sum of packet lengths to derive the features for the SVM. The cumulative sum is computed by adding the lengths of outgoing packets and subtracting the lengths of incoming packets. However, since the RBF kernel, in contrast to the aforementioned edit-distance based SVM kernel, expects feature vectors to have the same dimension, they interpolated 100 points from the cumulative sums. Furthermore, they prepend the total incoming and outgoing number of packets and bytes. As a result, they ended with 104 features to represent a traffic instance. Their evaluations demonstrate an attack success that ranges between 90% and 93% for 100 websites. It is worth mentioning that their dataset is the most realistic up to the date, including *inner* pages of sites that have spikes of popularity such as Google searches or Twitter links. Despite the high success rate of their attack, the authors conclude that the WF attack does not scale when applied in a real-world setting, as an adversary would need to train the classifier on a large fraction of all websites.

**k-Fingerprinting (k-FP) [13]:** Hayes and Danezis's k-FP attack is based on Random Forests (RF). Random Forests are ensembles of decision trees that are randomized and averaged so that they can generalize better than simple decision trees. Their feature sets include 175 features developed from features available in prior work, as well as timing features that had not been considered before, such as the number of packets per second. The random forest is not used to classify but as a way to transform these features into a different feature space: they use the leafs of the random forest to encode a new representation of the sites they intent to detect that is relative to all the other sites in their training set. Next, the new representation of the data is fed to a k-NN classifier for the actual classification. Their results show that this attack is as effective as CUMUL and achieves similar accuracy scores for the same number of sites.

All these attacks have selected their features mostly based on expertise and their technical knowledge on how Tor and the HTTP protocol work and interact with each other. As a result of manual feature engineering and standard feature selection, each proposed attack can be represented by a set of fingerprinting features. It is still unknown whether WF can be successfully deployed through automatic feature engineering based on implicit uninterpretable traffic features.

To the best of our knowledge, the only research that successfully applies deep learning to a similar problem is the network protocol recognition on encrypted traffic with a Stacked Denoising Autoencoder (SDAE) done by Wang [34]. His approach achieves a 90% recognition rate, which is a promising indicator for deep learning application to anonymized traffic.

The first effort to apply a DL-based approach to WF was made by Abe and Goto [1], where they evaluated a SDAE on the Wang-kNN's dataset. Their classifiers do not outperform the state-of-the-art, but nevertheless achieve a convincing 88% on a closed world of 100 classes. It is fair to assume that the lower performance is due to the lack of a sufficient amount of training data for a deep neural network, which, as we confirm later in our paper, is essential for the deep learning performance. Moreover, the work does not assess applicability of other deep learning algorithms to the problem. In this paper we explore three deep learning methods when applied to a significantly larger closed world of varying sizes, trained on sufficient amounts of data and evaluated in context of dynamic changes of web content over time. We provide a more extensive tuning of the DL-based attacks and finally achieve a similar accuracy to the state-of-the-art WF attacks.

### III. THREAT MODEL

In this paper we consider an adversary similar to the one considered in prior work in WF, namely a *passive* and *local* network-level adversary. Figure 1 shows an overview of this WF scenario. A passive adversary only records network packets transmitted during the communication and may not modify them or cause them to drop, and may not insert new packets into the stream of packets. A local adversary has a limited view of the network. In particular, in Tor, such an adversary typically owns the entry node to the Tor network (also known as *entry guard*), or has access to the link between the client and the entry. Examples of entities that have this level of visibility range from Internet Service Providers (ISP), Autonomous Systems (AS) or even local network administrators. Note that an adversary that owns the entry guard can decrypt the first layer of encryption and access Tor protocol messages. In this work, we assume an ISP-level adversary that collects traffic at the TCP layer and infers the cells from TCP packets [31]. Obviously, all work on WF assumes the adversary cannot decrypt the encryption provided by Tor, as message contents would immediately reveal the identity of the website.

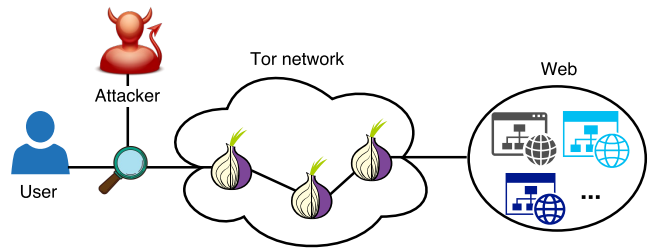

Fig. 1: The client visits a website over the Tor network. The adversary can observe the (encrypted) traffic between the client and the *entry* to the Tor network.

In the WF literature, it is common for the evaluation of the attack to assume a closed world of websites. This means that the user can only visit pages that the adversary has been able to train on. This assumption, commonly known as the *closed-world* assumption, has been deemed unrealistic [25] as the size of the Web is so large that an adversary can only train on a tiny fraction of the Web. For this reason, many studies have also evaluated the more realistic *open world*, where the user is allowed to visit pages that the adversary has not trained on. The closed world is still useful to compare existing attacks and defenses. In this study, we evaluated both the closed world and the open world.

## IV. DATA COLLECTION

One of the prerequisites for deep learning is an abundance of training data required to learn the underlying patterns. Processing sufficient amounts of representative data enables the deep neural network to not only precisely reveal the identifying features but also generalize better to unseen test instances. In prior work on WF in the context of Tor, the datasets that were collected are relatively limited in size, both in terms of classes (i.e. the number of unique websites) as well as instances (i.e. the number of traffic traces per website). To properly evaluate our proposed deep learning approach and explore how existing models can benefit from extra training data, we used a distributed setup to collect various new datasets that accommodate these requirements.

### A. Data collection methodology

For the data collection process, we used 15 virtual machines on our OpenStack-based private cloud environment. Each VM was provisioned with 4 CPUs and 4GB of RAM. To each VM, 16 worker threads were assigned, which each had their separate `tor` process (version 0.2.8.11). Page-visit tasks, consisting of starting the Tor browser (version 6.5) and loading the target web page, were then distributed among the 240 concurrent worker threads. Web pages were given 285 seconds to load, before the browser was killed and the visit marked as invalid. Upon loading the page, it was left open for an additional 10 seconds, after which the browser was closed and any profile information was removed.

By leveraging network namespaces and `tcpdump`, we isolated and captured the traffic of each `tor` process. Due to storage constraints, and since the packet payloads are encrypted and thus do not have value for the adversary, we extract metadata from the traffic trace and discard the encrypted payload. More precisely, we capture (1) the timing information, (2) the direction and (3) the size of the TCP packet. We follow the approach proposed by Wang and Goldberg [31] to extract Tor cells from the captured TCP packets. Our final representation of the traffic trace is a sequence of cells, where each cell is encoded as 1 when transmitted from the client to the website and as $-1$ when captured in the opposite direction. For the purpose of sanity checks and validation, information on the Tor circuit that was used for the page visit is also recorded.

It should be noted that, in contrast to prior work [31], the Tor entry guard node was not pinned over the course of our experiments. The reason for this is twofold. First, compared to prior data collection, we use significantly more concurrent processes. If the same entry guard would be used by the 240 browser instances, this could overload the entry guard, possibly affecting the network traces. Second, by using a variety of entry guards, the trained models are agnostic to the intrinsics of a specific entry guard. This means that the model of the adversary is not only applicable in a targeted attack on a single victim, but can be launched against any Tor user.

### B. Datasets

Since the WF adversary's goals might vary widely and as there are no statistics about which pages Tor users browse to, there can be no definitive set of sensitive websites for WF research. Moreover, since we aim to compare various approaches with each other, the actual choice of websites is not essential as long as it is consistent. The list of websites we chose for our evaluation comes from the Alexa Top Sites service, the source widely used in prior research on Tor.

In total, we evaluate our deep learning approach in comparison with traditional methods on three different datasets. This section details how these datasets were chosen and obtained.

*1) Closed world:* For the dataset under the *closed world* assumption, we collected up to 3,000 network traces for visits to the homepage of the 1,200 most popular websites according to Alexa. The list of popular websites was first filtered to remove duplicate entries that only differ in the TLD, e.g. in the case of `google.com` and `google.de`, only the former was included in the list. Data for these 1,200 websites was collected in four iterations, consisting of 300 websites each. An iteration was again split up into 30 batches, with each batch performing 100 network traces per websites. After each batch, the 240 `tor` processes were restarted and data directories were removed, forcing new circuits to be built with (new) randomly selected entry guards. Network traces for each of the four iterations were collected over approximately 14 days per group, starting from January 2017.

After collecting data on the 3.6 million page visits, we filtered out invalid entries, which were due to a timeout, or a crash of the browser or Selenium driver. Websites with a high amount of invalid page visits were removed from our dataset. Additionally, using the similarity hash of the web page's HTML content [7] and the perceptual hash of the screenshot [3], we detected and excluded websites with exactly the same content. Moreover, we filtered out websites that had no content, denied all requests coming from Tor, or showed a CAPTCHA for every visit. Finally, we balanced the dataset to ensure the uniform distribution of instances across different sites by fixing the same number of traces for every site. After this filtering process, our biggest closed world dataset consists of 900 websites, with 2,500 valid network traces each. In the remainder of the text, we refer to this dataset as $CW_{900}$. Similarly, for datasets that are composed of a subset of this one we use a corresponding representation: the datasets for the top 100, 200 and 500 websites are referred to as $CW_{100}$, $CW_{200}$ and $CW_{500}$ accordingly.

*2) Revisit over time:* For the top 200 websites, we obtained additional periodic measurements. More precisely, for these websites we collected 100 test network traces per website 3 days, 10 days, 4 weeks, 6 weeks and 8 weeks after the end of the initial data collection for these 200 websites. Each test set is collected within one day. As a result, our *revisit-over-time* dataset provides 500 network traces for each of the top 200 websites collected over a 2-month period ($CW_{200}$ was collected over 2 weeks).

*3) Open world:* Since the *open world* data is only used for testing purposes (which differs from some of the open world evaluations), we collected only a single instance for each page in the open world. In total, we collected network traces for the top 400,000 of Alexa websites.

We collected additional 2,000 test traces for each website of the monitored closed world $CW_{200}$ (400,000 instances in total). As a result, we conduct the open world evaluation on 800,000 test traffic traces, half from the closed world and half from the open world (a 4-fold increase compared to the largest dataset considered in prior work [13], [24]). We provide the motivation for this experimental setting in Section V-B5.

### C. Ethical considerations & data access

For our data collection experiments, we performed around 4 million page visits over Tor. It is highly unlikely that this had any impact on the top websites, which each receive multiple millions of requests every day. We consider the impact on the Tor network to be limited as well: The Tor Project estimates that during the time we performed our experiments, approximately 2 million clients were concurrently connected to the Tor network. As such, the 240 clients we used are only a minor fraction of the total number of active clients. Furthermore, we made the data publicly available upon acceptance of this paper, allowing other researchers to evaluate other approaches without having to collect new data samples.

## V. EVALUATION

In this section, we conduct a reevaluation of the state-of-the-art WF methods discussed in the related work of Section II to confirm their reproducibility on our dataset. We then evaluate the proposed attacks based on the three chosen deep learning (DL) algorithms and compare them to the previously known techniques.

### A. Reevaluation of state-of-the-art

We aim to enable a systematic comparison between our work and that of Wang et al. [30], Panchenko et al. [24] and Hayes et al. [13], not only to guarantee a fair assessment by evaluating on new data, but also to analyze (1) the practical feasibility of the attack on a significantly larger set of websites, (2) the impact of collecting more instances or traces per website on the classification accuracy, and (3) the resilience of trained models to concept drift with a growing time gap between training and testing.

The goal of the first closed world experiment is to confirm whether we can reproduce the three WF attacks of prior work [30], [24], [13] and to assert whether we obtain similar classification results as those reported by the respective authors, but on a different training and testing dataset similar in size. We reuse the original implementation of the authors to carry out the feature extraction and subsequently execute the training and testing steps. All results reported in this section are computed via 10-fold cross-validation.

The following results were obtained on a Dell PowerEdge R620 server with 2x Intel Xeon E5-2650 CPUs, 64GB of memory and 8 cores on each CPU with hyperthreading, resulting in 32 cores in total each running at 2GHz. Wang's $k$-NN based attack ran on a single core as the stochastic gradient descent method to find the best weights for $k$-NN classification could not be parallelized without sacrificing some classification accuracy. Panchenko's CUMUL attack trains an SVM model which requires a grid search to find the best $C$ and $\gamma$ parameter combination for the RBF kernel. As the native libSVM library is not multi-core enabled, the parameter combination tests ran as parallel processes each on a single core, with the time reported being the one of the slowest $C$ and $\gamma$ parameter combination test.

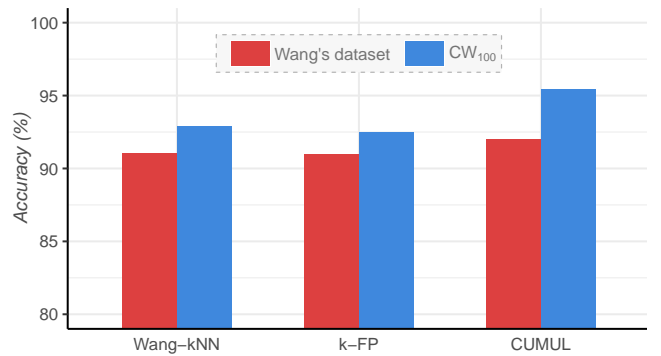

Fig. 2: Re-evaluation of traditional WF attacks on new data

Figure 2 shows the closed world classification accuracy obtained through cross-fold validation for the three traditional WF attacks on a $CW_{100}$ dataset with 100 traces per website. For the same set of website instances, the $k$-NN algorithm of Wang et al. reports a classification accuracy of 92.87% on our new data set, whereas the CUMUL algorithm of Panchenko et al. and the $k$-FP attack by Hayes et al. respectively report accuracy results of 95.43% and 92.47%. The obtained results are in line with those originally reported by the authors themselves albeit on other data sets. For this particular setup, the CUMUL WF attack turned out to be the most accurate.

In the second experiment, we evaluate the same traditional methods on 100 websites, but with a growing number of traces per website, to investigate whether the classification accuracy improves significantly when provided with more training data and whether one WF attack method is consistently better than another.

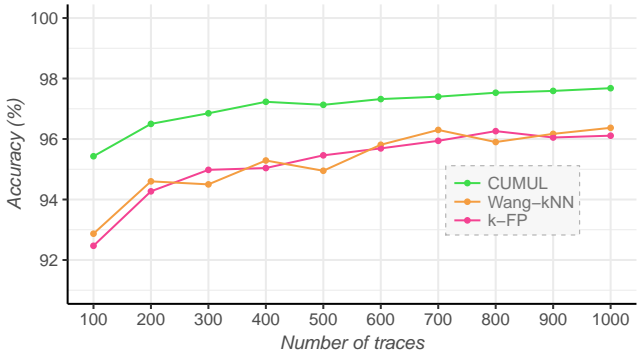

Fig. 3: Impact on the classification accuracy for a growing number of website traces

In Figure 3, we depict the classification accuracy in a closed world experiment where the number of website instances grows from 100 to 1,000 traces. Our results show that the CUMUL attack consistently outperforms the two other methods. For all methods, the improvement becomes less evident after about 300 website traces. Another interesting observation is that each WF attack − when given sufficient training data − converges to a classification accuracy of approximately 96-97%. However, we experienced scalability issues with the $k$-NN based attack by Wang et al., given that the classification running times were at least an order of magnitude higher than those of the CUMUL and $k$-FP attacks.

In a third experiment, we assess how the classification accuracy drops when the number of websites increases for a fixed amount of training instances. Given that the CUMUL attack consistently outperformed the other two methods on our dataset, and was superior in resource consumption, we only report the results for CUMUL. We reevaluate the CUMUL classifier on our closed worlds $CW_{100}$, $CW_{200}$, $CW_{500}$ and $CW_{900}$ with a fixed number of traffic traces: 300 per website.

Table I illustrates that the CUMUL attack obtains a reasonable 92.73% 10-fold cross-validation accuracy for 900 websites using 300 instances each, and a parameter combination of $log_2(C) = 21$ and $log_2(\gamma) = 5$. In general, we observe that the performance degrades gradually with a growing size of the closed world. Moreover, doubling the initial amount of instances gives an advantage of up to 2%, while the amounts higher than 300 stop providing any significant improvement. The biggest weakness is that for each experiment one must execute the grid search to ensure the best classification results, and certain parameter combination tests take a long time to converge with no guarantee of a gain in accuracy.

TABLE I: CUMUL accuracy for a growing closed world (with 100 traces per website, 300 traces, and the best achieved accuracy for a varying number of traces).

| Dataset | CUMUL (100tr) | CUMUL (300tr) | CUMUL (best) |
|---|---|---|---|
| $CW_{100}$ | 95.43% | 96.85% | 97.68% (2000tr) |
| $CW_{200}$ | 93.58% | 95.93% | 97.07% (2000tr) |
| $CW_{500}$ | 92.30% | 94.22% | 95.73% (1000tr) |
| $CW_{900}$ | 89.82% | 92.73% | 92.73% (300tr) |

TABLE II: Time required to find optimal RBF parameter values for $C$ and $\gamma$ for SVM based classification.

| Traces | $CW_{100}$ | $CW_{200}$ | $CW_{500}$ | $CW_{900}$ |
|---|---|---|---|---|
| 100 | 3 min | 8 min | 139 min | 771 min |
| 200 | 10 min | 48 min | 684 min | 3027 min |
| 300 | 19 min | 99 min | 1230 min | 4031 min |
| 400 | 29 min | 134 min | 1490 min | > 6000† min |
| 500 | 34 min | 169 min | 1541 min | > 6000† min |
| 1000 | 41 min | 844 min | 5016 min | > 6000† min |
| 2000 | 41 min | 844 min | 5016 min | > 6000† min |

†Aborted experiments.

Table II gives an overview of the running times (in minutes) to find the best $C$ and $\gamma$ parameter values for the RBF kernel. We aborted those experiments where the grid search took more than four days to complete. While there is a trend of increasing values for these parameters with a growing number of websites and instances, we could not find a strong correlation that would enable us to eliminate the grid search altogether.

As a result, we choose CUMUL as the reference point for comparing our proposed method with the state-of-the art. This decision is driven by the fact that CUMUL performed the best on our closed worlds, and proved to be more practically feasible. We acknowledge that the k-FP attack has the potential to work better in our open world evaluation. However, over the course of our scalability experiments, k-FP did not scale to 50,000 training instances. The experiment consumed more than 64GB memory and took longer than the allocated 4 days, and thus was aborted. With our open world datasets consisting of 800,000 instances (and 400,000 training instances), such high resource consumption demands strongly limit large scale evaluation. CUMUL on the other hand scales up to 400,000 training instances. Therefore, we further evaluate our DL-based approach in comparison to CUMUL, which outperformed the other traditional WF techniques and which was practically feasible on a larger scale.

*B. Deep Learning for Website Fingerprinting*

Here we provide a detailed outline of our DL-based methodology. DL provides a broad set of powerful machine learning techniques with deep architectures. Deep neural networks (DNN), which underlie DL, exploit many layers of non-linear mathematical data transformations for automatic hierarchical feature extraction and selection. DNN demonstrate a superior ability of feature learning for solving a wide variety of tasks. In this study we apply three major types of DNNs to WF: a feedforward SDAE, a convolutional CNN and a recurrent LSTM.

*1) Problem definition:* In our proposed method, we follow prior work and formulate WF as a classification problem. Namely, we perform a supervised multinomial classification, where we train a classifier on a set of labeled instances and test the classifier by assigning a label out of a set of multiple possible labels to each unlabeled instance. In WF, a traffic trace $t$ captured from a single visit to a website is an instance of the form $(\mathbf{f_t}, c_t)$, where $\mathbf{f_t}$ is the feature vector of the traffic trace and $c_t$ is the class label that corresponds to the website that generated this traffic. Assuming a closed world of $N$ possible websites, label $c_t$ belongs to the set $\{0, 1, \ldots, N-1\}$. As such, we state the WF problem as follows: assign a class label to each anonymous traffic trace in a dataset based on its features.

The classifiers used in related work successfully solved this problem by carefully constructing feature vectors, as described in Section II. Our proposed classifier, based on a DNN, integrates feature learning within the training process, enabling it to classify traffic traces simply based on their initial representation. Thus, for a DL classifier, the form of the input instance changes to $(\mathbf{r_t}, c_t)$, where $\mathbf{r_t}$ is a raw representation of a traffic trace that can be interpreted by a neural network.

In essence, we represent a traffic trace as a sequence of successive Tor cells that form the communication between the target user and the visited website. As a result, an input instance of our DNN-based classifier is a series of $1$ and $-1$ of variable length, based on which model performs feature learning and website recognition. Our choice of this format is also supported by the fact that neural networks generally work with real numbers from the compact interval $[-1, 1]$ due to the nature of the mathematical operations they perform. Moreover, by providing the input data in such a format, we avoid having to rescale and/or normalize the values and thus mitigate a possible information loss coupled with the preprocessing step.

Out of all existing types of DNNs and corresponding DL algorithms, we evaluate three major types of neural networks: feedforward, convolutional and recurrent. We choose to apply the models that provide the capabilities and architectural characteristics to perform the task of automated feature extraction and to benefit from the nature of our input data. We refer to the Appendix for a more elaborate and in-depth discussion on the DL algorithms, which we consider to be conceptually the most well-suited for the WF task at hand.

The first DNN we apply is a classifier called Stacked Denoising Autoencoder (SDAE) – a deep feedforward neural network composed of Denoising Autoencoders (DAE). An Autoencoder (AE) is a feedforward network specifically designed for feature learning through dimensionality reduction. Stacking multiple AEs as building blocks to form a deep model allows for hierarchical extraction of the most salient features of the input data and performing classification based on the derived features, which makes SDAE a promising model for our WF problem.

The next proposed DNN is a Convolutional Neural Network (CNN) – a classifier built on a series of convolutional layers. Convolutional layers are also used for feature extraction, starting with low-level features at the first layer and building up to more abstract concepts going deeper in the network. CNN's methodology for achieving that differs from that of SDAE. Convolutional layers learn numerous filters that reveal regions in the input data containing specific characteristics. These input instances are then downsampled with the special regions preserved. In such a way the CNN searches for the most important features to base the classification on. Furthermore, while SDAE has to be pretrained block by block, CNN requires minimum preprocessing.

The final chosen DNN is yet another type of a neural network, very different in its fundamental properties from the first two. A classifier called Long-Short Term Memory network (LSTM) is a special type of a recurrent neural network that has enhanced memorization capabilities. Its design allows for learning long-term dependencies in data, enabling the classifier to interpret time series. Our input traffic traces are essentially time series of Tor cells, and temporal dynamics in these series are expected to be highly revealing of the contained website fingerprint, thus the choice of the model.

We used `Keras`[10] with `Theano`[28] backend for the implementation of the DNN classifiers. The source code is publicly available on the following webpage: https://distrinet.cs.kuleuven.be/software/tor-wf-dl/.

*2) Hyperparameter tuning and model selection:* The adversary has to empirically select a DNN model to apply for WF. For that, the adversary should tune the hyperparameters of the DNN to achieve the best classification performance and, at the same time, enhance its capabilities to generalize well to unseen traffic traces.

Performing an automatic search of the best hyperparameters – be that an exhaustive grid search, a random search or another search algorithm – is highly effective but computationally expensive at the same time. In our work, we evaluate the DL algorithms applied to WF by performing semi-automatic hyperparameter tuning, where we exploit the knowledge of each hyperparameter's impact. Namely, the main strategy is as follows:

- The adversary chooses a *representative subsample* of the given dataset and splits it *randomly* into training set, validation set and test set in the following proportion: 90% - 5% - 5%
- Next, the adversary defines the limits of the model capacity based on the amount of available training data. On the one hand, the model has to be expressed with a sufficient amount of parameters in order to be able to learn the problem. On the other hand, there has to be *much fewer* trainable parameters than available training instances in order to avoid overfitting. The model's capacity is defined through its structure and hyperparameters, different for each DNN. The adversary has to define the search spaces for each hyperparameter.

- In our evaluation a special form of Bayesian optimization is applied for hyperparameter tuning, specifically a Tree of Parzen Estimators (TPE)[2] implemented in `hyperopt` library. Through this algorithm the adversary automates the tuning process within previously defined search spaces.
- The optimization algorithm returns the best combination of values and the network structure based on the test results. If the adversary finds the model's test performance satisfactory, he selects this model. Otherwise, he adjusts the search spaces and repeats the tuning procedure.
- Finally, the adversary builds and initializes the selected learning model and applies it to the *whole* dataset to deploy the actual WF attack.

Traditional machine learning methods used for WF in the related work (such as SVM, k-NN and RF, as presented in Section II) also require hyperparameter tuning, but on a smaller scale than DL. Nevertheless, tuning the parameters of the DL model becomes even more feasible in comparison to traditional models due to the parallelism of DL algorithms. As learning algorithms of neural networks are inherently parallel, graphical processing units (GPUs) can take advantage of this characteristic. Performing hyperparameter tuning on GPUs compromises for intense computational requirements allows for rapid feedback of the model. For our DL experiments we use two Nvidia GeForce GTX 1080 GPUs with 8GB memory and 2560 cores each and one TITAN Xp with 12GB memory and 3840 cores to accommodate parallelized training of the DNNs. The training runtime reported in this paper should therefore be interpreted in association with said platforms.

Table III includes the list and the values of the hyperparameters we tuned, together with the corresponding intervals within which we vary the values. Each hyperparameter controls a certain aspect of the DL algorithm: architecture (structural complexity of the network), learning (the training process) and regularization (constraint of the learning capabilities applied order to avoid *overfitting*, which occurs when the model memorizes the training data instead of learning from it). Note that in order to reduce the search space, we limited our models to the same learning and regularization parameters for each network layer.

The adversary is supposed to select the DL-based model *once* given a sample crawled for a desired closed world of websites. Similarly, we perform the model selection on the $CW_{100}$ dataset, as defined in Section IV, in order to limit the computational requirements. Given a proper tuning procedure and a sufficiently large amount of training instances for each class, the chosen model is expected to learn the problem (learn to extract the fingerprints), and at the same time generalize well to the other closed world datasets. In fact, the adversary capable of crawling large amounts of data can compensate on hyperparameter tuning.

The final selected models of SDAE, CNN and LSTM used for evaluation are described in Table III. The amount of LSTM units has to be adjusted for the bigger closed worlds to increase expressive capacity. Note that due to the LSTM's backpropagation through time constraints, we have to trim the traffic traces to the first 150 Tor cells (we elaborate on the reason for that in Appendix).

Further in this subsection we present the experimental results of the DL-based WF attack on the crawled dataset. Namely, we evaluate the three chosen DNNs on the closed worlds of various sizes and on the open world. We also assess their generalization capabilities by testing their resilience to concept drift on data periodically collected over 2 months. Furthermore, we compare results to CUMUL, being the most accurate traditional WF method.

*3) Closed world evaluation:* In this study, we evaluate the SDAE, CNN and LSTM networks on four closed worlds of different sizes, namely $CW_{100}$, $CW_{200}$, $CW_{500}$ and $CW_{900}$. We use the models selected by performing hyperparameter tuning on the $CW_{100}$ dataset, according to the aforementioned methodology. To ensure the reliability of our experiments, we estimate the models' performance by conducting a 10-fold cross-validation on each dataset. We use two performance metrics to evaluate and compare the models with each other: the test *accuracy* (classification success rate, which needs to be maximized) and the test *loss* (a cost function that reflects the significance of classification errors made by the model, namely the categorical cross-entropy, that needs to be minimized, as explained in the Appendix).

The aspect that had the greatest impact on the performance over the course of our experiments was the amount of training data (i.e. the amount of traffic traces for each website), which is in line with our expectations and justifies the extensive data collection. Indeed, for every closed world experiment, we observed significant improvements for a growing amount of traces. One example of this trend is given in Figure 4 for the $CW_{100}$ dataset, where we vary the amount of instances from 100 to all available 2,500 per class. The Table IV reports on the actual metrics' values and the corresponding runtimes.

First and foremost, from these results we can confirm the *feasibility of the WF attack based on a DL approach with automatic feature learning*. We observe how classification accuracy and loss function gradually improve for all models, in the end reaching the 95.46, 96.66 and 94.02% success rate for SDAE, CNN and LSTM model accordingly. These results are comparable to the ones achieved by traditional approaches in Section V-A.

If we compare the three DNNs with each other, we observe that the SDAE and CNN networks consistently perform better than the LSTM in terms of classification accuracy, with CNN being the most performant. Nevertheless, knowing that the LSTM classifies traffic traces based solely on their first 150 Tor cells (compared to the SDAE and CNN that use up to 5,000 and 3,000 cells from each trace), the achieved performance still appears promising. Our interpretation is that even a small part of the traffic trace is sufficient for website recognition up to 94% accuracy when deploying a model that is able to exploit temporal dependencies of the input sequence. Notably, LSTM performs much poorer when trained on fewer traffic traces than SDAE and CNN, but later gains comparable recognition rate at 1000 training instances per class.

TABLE III: Tuned hyperparameters of the selected DL models.

|  | SDAE | | CNN | | LSTM | |
|---|---|---|---|---|---|---|
| **Hyperparameter** | **Value** | **Space** | **Value** | **Space** | **Value** | **Space** |
| optimizer | SGD | SGD, Adam RMSProp | RMSProp | SGD, Adam RMSProp | RMSProp | SGD, Adam RMSProp |
| learning rate | 0.001 | 0.0001 .. 0.1 | 0.0011 | 0.0009 .. 0.0025 | 0.001 | 0.0001 .. 0.1 |
| decay | 0.0 | 0.0 .. 0.9 | 0.0 | 0.0 .. 0.9 | 0.0 | 0.0 .. 0.9 |
| batch size | 32 | 8 .. 256 | 256 | 8 .. 256 | 128 | 32 .. 256 |
| training epochs | $\leq 30$ | 1 .. 100 | 3-6 | 1 .. 20 | $\leq 50$ | 1 .. 100 |
| number of layers | 5 | 3 .. 7 | 8 | 6 .. 10 | 4 | 3 .. 6 |
| input units | 5000 | 200 .. 5000 | 3000 | 200 .. 5000 | 150 | 70 .. 1000 |
| hidden layers units | 1000, 500, 300 | 200 .. 3000 | — | — | 64, 64 / 128, 128 | 64 .. 256 |
| dropout | 0.1 | 0.0 .. 0.5 | 0.1 | 0.0 .. 0.5 | 0.22 | 0.0 .. 0.5 |
| activation | tanh | tanh, sigmoid, relu | relu | tanh, relu | tanh | tanh, sigmoid, relu |
| pretraining optimizer | SGD | SGD, Adam | — | — | — | — |
| pretraining learning rate | 0.1 | 0.01 .. 0.1 | — | — | — | — |
| kernels | — | — | 32 | 4 .. 128 | — | — |
| kernel size | — | — | 5 | 2 .. 50 | — | — |
| pool size | — | — | 4 | 2 .. 16 | — | — |

TABLE IV: Accuracy, loss and runtime of the DL models (SDAE, CNN, LSTM) for $CW_{100}$ and a growing number of traces.

|  | SDAE | | | CNN | | | LSTM | | |
|---|---|---|---|---|---|---|---|---|---|
| **Traces** | **Accuracy** | **Loss** | **Runtime** | **Accuracy** | **Loss** | **Runtime** | **Accuracy** | **Loss** | **Runtime** |
| 100 | 85.00% | 0.5902 | 0 min | 81.25% | 0.8276 | 0 min | 40.60% | 2.2132 | 9 min |
| 200 | 87.30% | 0.5252 | 1 min | 86.63% | 0.5793 | 0.5 min | 57.30% | 1.5471 | 17 min |
| 500 | 91.34% | 0.3576 | 1 min | 91.43% | 0.3877 | 1 min | 79.54% | 0.7848 | 40 min |
| 1000 | 92.64% | 0.2950 | 2 min | 94.72% | 0.2545 | 1.5 min | 91.63% | 0.3555 | 63 min |
| 1500 | 94.49% | 0.2314 | 4 min | 95.95% | 0.1855 | 2 min | 91.93% | 0.3055 | 66 min |
| 2000 | 95.17% | 0.1955 | 6 min | 96.14% | 0.1699 | 3 min | 93.98% | 0.3277 | 67 min |
| 2500 | 95.46% | 0.1968 | 7 min | 96.26% | 0.1784 | 5 min | 94.02% | 0.3204 | 76 min |

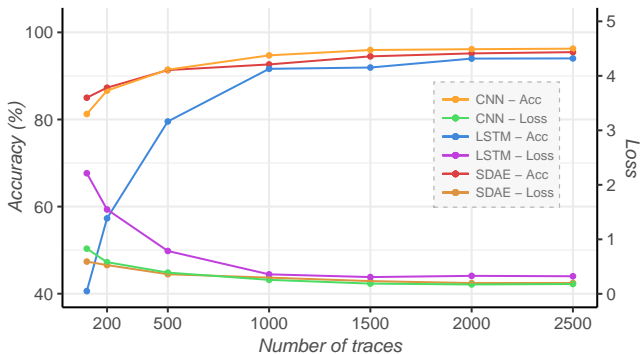

Fig. 4: Accuracy, loss and evaluation time of the DL models (SDAE, CNN, LSTM) for $CW_{100}$ and a growing number of traces

Next, we assess whether the selected DL models tuned on $CW_{100}$ perform similarly when applied to the larger datasets: $CW_{200}$, $CW_{500}$ and $CW_{900}$. The results of the DL-based WF for all closed world datasets are presented in Table V, expressed in classification accuracy, loss function and runtime. The time reported in the table is the average time required to build, train and evaluate a model. We observe that for larger closed worlds the performance of the three DL models gradually decreases following a similar trend. The closed world evaluation results remain comparable to CUMUL's results presented in Table I in the previous subsection. Figure 5 compares the DL-based methods to CUMUL. This comparison illustrates that our DL-based attack can indeed successfully learn the fingerprinting features in an automated manner. Furthermore, the training method itself is highly parallelizable on GPU hardware resulting in a faster and therefore more practical closed world WF attack.

The presented experiments on the closed world reflect the model's ability to classify traffic traces that are collected at the same moment as the training data. Even though we prove that such a WF attack is possible, we do not address the question of eliciting the concrete data features that the models take decisions upon. In other words, just based on this experiment, we cannot certainly infer if the DNN reveals the actual website fingerprint for deanonymization, or also learns occasional dynamics in the traffic data instead that just happens to enable recognition. The next experiment is intended to reveal how well our DNNs are able to extract the fingerprint and generalize to new data.

*4) Concept drift evaluation:* The challenge of recognizing traffic traces collected over time was first addressed by Juarez

TABLE V: Accuracy, loss and runtime of the DL models (SDAE, CNN, LSTM) for each closed world and 2,500 traces.

| Dataset | SDAE | | | CNN | | | LSTM | | |
|---|---|---|---|---|---|---|---|---|---|
| | Accuracy | Loss | Runtime | Accuracy | Loss | Runtime | Accuracy | Loss | Runtime |
| $CW_{100}$ | 95.46% | 0.1968 | 7 min | **96.66%** | 0.1699 | 5 min | 94.02% | 0.3204 | 76 min. |
| $CW_{200}$ | 95.76% | 0.1822 | 14 min | **96.52%** | 0.1774 | 8 min | 93.10% | 0.3292 | 91 min |
| $CW_{500}$ | **95.04%** | 0.2243 | 34 min | 92.31% | 0.3732 | 12 min | 90.80% | 0.3163 | 257 min |
| $CW_{900}$ | **94.25%** | 0.2530 | 52 min | 91.79% | 0.4278 | 20 min | 88.04% | 0.3601 | 276 min |

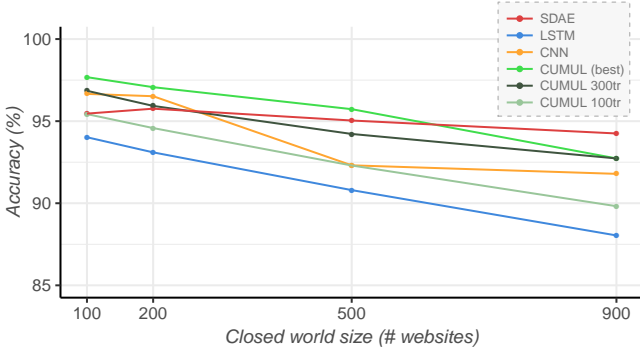

Fig. 5: DL (SDAE, CNN, LSTM) vs. CUMUL for a growing size of the closed world from 100 to 900 websites.

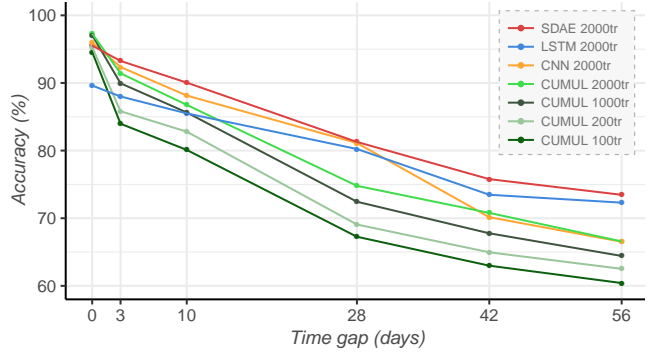

Fig. 6: DL (SDAE, CNN, LSTM) vs. CUMUL resilience to concept drift: evaluation of $CW_{200}$ over time.

et al. [19]. They showed that classification accuracy drops drastically when testing the model on traffic captured 10 days after training. This time effect is explained by constant content changes of the websites, which of course may affect the identifying fingerprints. Another possible reason for the performance drop is that the classifier trained and evaluated at one moment in time might overlook the stable fingerprint and learn the temporary features instead. In general such an occurrence is known as *concept drift* – a change over time in the statistical properties of the class that the model is trying to predict. Therefore, the recognition might become less accurate over time. A model resilient against concept drift is the one that manages to capture the salient traffic features maximally correlated with the website fingerprint and thus remains performant over time. To reveal if our DNNs detect the actual website fingerprints and assess how well they perform in case of traffic changes, we train the models on a closed world and test them on data collected from visiting websites of the same closed world periodically over 2 months. In order to fairly compare DL-based methods to CUMUL, we have to evaluate them on the same dataset with the same amount of traces. Due to CUMUL's scalability issue, the biggest dataset possible to use for this evaluation is $CW_{200}$ with 2,000 training instances. Even though this is not the largest dataset we collected, it is still twice bigger than the closed worlds normally used in prior works. Thus we train models on the whole $CW_{200}$ dataset (with 2,000 training traces) and test them on the *revisit-over-time* dataset (as defined in Section IV).

The results are depicted in Figure 6 for DL and traditional CUMUL. The plot indicates the WF performance of various models trained on $CW_{200}$ and evaluated on traffic re-collected 3 days, 10 days, 4 weeks, 6 weeks and 8 weeks after training.

The figure demonstrates how the classification accuracy decreases and the classification loss increases gradually and drastically over time. These results illustrate the high generalizing abilities of both the evaluated models. Despite a significant 2-month time gap between the moment of training and the last evaluation, the DL algorithms are still capable to correctly deanonymize at least 66% out of 2,000 website visits. We witness a rather small accuracy drop in the first 3 and 10 days for all three DL models, which may be acceptable for an adversary who would prefer to use the built WF classifier for several more days rather than repeat the data collection and training process every day. In total, SDAE loses 22% of accuracy over 2 months, CNN loses 29%, while LSTM only loses 17%. Notably, being the most performant DL model on the day of training, CNN generalized worse than SDAE or LSTM. Despite the fact that the LSTM model (which still makes decision just based on the first 150 cells in the input sequence) is initially outperformed by both SDAE and CNN, after one month its accuracy catches up with that of the SDAE. Moreover, after 1 month the LSTM loss values are lower than those of the SDAE, which means that even though the LSTM outputs less correct predictions, it is overall more certain of these predictions. This obviously speaks in favor of LSTM's high generalization abilities, in line with our best expectations.

Our SDAE and CNN approaches outperform CUMUL with up to 7% over the course of 2 months. In total CUMUL loses 31%. LSTM network starts outperforming CUMUL after approximately 2 weeks. As such, this comparison not only shows that our approach indeed automates the feature engineering, but also that the learned implicit features (hidden in the neural network) are more robust against website changes over time. Notably, CUMUL is found to significantly improve its generalization abilities when trained on larger amounts of traffic traces per website, which proves that DL-based classifiers are not alone in their requirement for a bigger training data for the highest performance.

The main conclusion here is that the DL-based classifiers are capable of extracting stable identifying information from the closed world traffic which allows for its deanonymization with a high success rate, even several days after training.

*5) Open world evaluation:* This study compares DL-based WF attacks and CUMUL for the open world evaluation. The goal is to assess the classifier's ability to distinguish a traffic trace generated by a visit to one of the monitored websites from a traffic trace generated by a visit to any other unknown website. Our methodology for the open world evaluation differs from prior work in several aspects. We aim to provide a fair comparison of the classifiers by reducing possible bias. To this purpose we have to depart from the realistic WF setting and adapt the following assumptions:

- We model the monitored websites by training the classifier solely on the traffic traces of the websites an adversary is aiming to detect. By doing so, we assess the abilities of the learning algorithms to distinguish seen and unseen websites. In previous studies on WF, it has been argued that an adversary may improve the attack by additionally collecting and training on traffic of known websites that he is not interested in identifying, which is of course a possibility given sufficient resources. But here we do not provide any helping patterns of the open Web to the classifiers to not distort their actual performance.
- We test the classifiers on balanced datasets: monitored and unknown websites in proportion 50%-50% (meaning that random classification would be accurate on average 50% of a time). Thus, we do not attempt to infer the realistic ratio, especially knowing that modeling an open world of a realistic scale poses large issues: (1) the effect of the hypothesis space complexity, as shown by Panchenko et al. [24], and (2) the *base rate fallacy*, demonstrated by Juarez et al. [19]: even a highly accurate classifier trained on the monitored websites with a very low prior probabilities of visit cannot be fully confident of its predictions. Instead we assume a standard uniform probability distribution of visits to the monitored and unknown sets. With such evaluation the classifier's errors are more prominent and allow for a clearer comparison.
- Following the earlier reasoning, we use Alexa websites for both, monitored and unknown sets. Choosing a particular set of monitored websites characterized by patterns that are not common to the whole Web would introduce classification bias with unpredictable impact on comparison. In order to objectively compare the studied classifiers, we demonstrate their abilities to distinguish seen and unseen fingerprints belonging to the websites of the same category (in our case, most popular websites).

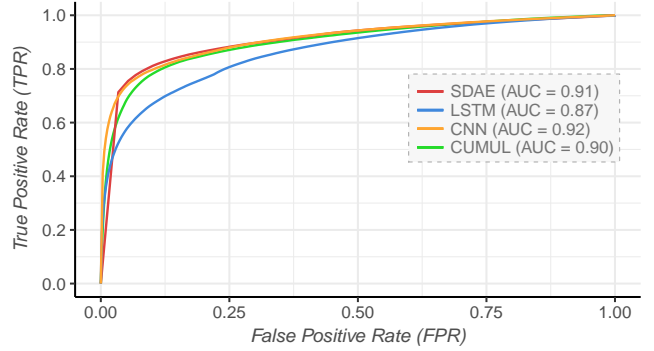

Fig. 7: DL (SDAE, CNN and LSTM) vs. CUMUL in the open world setting for a monitored set of $CW_{200}$.

We evaluate the open world WF attack for an adversary who monitors a set of 200 websites, while the target user may visit 400,000 more unknown websites. As a result, our open world dataset consists of 800,000 visits through Tor: one-time visits to 400,000 various websites in the Web and 400,000 visits to the monitored $CW_{200}$. We train the models solely on 2,000 instances of $CW_{200}$ (thus obtaining the classifiers identical to those used for the closed world evaluation). Recall that earlier in the closed world evaluation section we already assessed their multinomial classification performance; the reported success rates indicate the ability of the classifiers to identify the exact visited monitored website. In this section we perform binary classification by testing the same models on our open world dataset. With this experiment we assess the classifiers' ability to recognize the input instance as a visit to a monitored or an unknown, earlier unseen website. The classifier makes decisions based on the cross-entropy loss function, which reflects its confidence in made predictions (Appendix elaborates on the cross-entropy as a measure of classification confidence). If the loss value is low enough, the adversary assumes that the classified website visit belongs to a set of monitored websites. If the entropy is bigger than a certain *confidence threshold*, the adversary decides to not trust the classifier's class prediction and concludes that the tested traffic trace was generated by an unknown website, thus causing the prediction uncertainty. By varying the confidence threshold, the adversary balances the True Positive and False Positive Rate according to their priorities.

In our evaluation, we plot the ROC curve for the three DL classifiers in order to define the optimal confidence threshold which separates the monitored websites traffic from unknown websites traffic. Both CNN and SDAE again outperform CUMUL, if only slightly, as demonstrated by Area Under Curve values in the same figure. The ROC curves for SDAE, CNN and LSTM are depicted in Figure 7 and demonstrate the relative performance of the suggested open world WF DL-based attacks within 200 monitored and 400,000 unknown websites. We observe that the CNN model performs better than SDAE, and both perform significantly better than the LSTM model. However, the adversary may improve the models by using the open world traces for validation during hyperparameter tuning . LSTM classifier is outperformed by two other DL models because it only processes the first 150 Tor cells, opposed to 5,000 by SDAE and 3,000 by CNN.

According to the ROC curves, an adversary may optimize the confidence threshold depending on their priority. For 200 classes, the categorical cross-entropy $E$ varies between 0 (absolute confidence of the classifier's prediction) to 5.3 (absolute uncertainty). The optimization examples are given in Table VI, where reduced thresholds allow to decrease FPR.

TABLE VI: DL vs. CUMUL in the open world setting.

| Model | Optimized for TPR | | | Optimized for FPR | | |
|---|---|---|---|---|---|---|
|  | $E$ | TPR | FPR | $E$ | TPR | FPR |
| SDAE | 0.005 | 80.25% | 9.11% | 0.001 | 71.30% | 3.40% |
| CNN | 0.033 | 80.11% | 10.53% | 0.013 | 70.94% | 3.82% |
| LSTM | 0.062 | 76.19% | 19.78% | 0.010 | 53.39% | 3.67% |
| CUMUL | 0.048 | 78.00% | 9.89% | 0.018 | 62.57% | 3.58% |

Our open world evaluation considers a large set of unknown sites in which the adversary cannot train, allowing us to test the generalization of our models in a large sample of the Web. Similarly to the state-of-the-art, we observe how our DL-based approach withstands a challenging open world scenario, providing high accuracy on the largest set of unknown sites.

In the previous subsections, we have shown the relative performance of various DL models in comparison with each other and with the traditional CUMUL classifier. In certain experimental settings we improved beyond the state-of-the-art, e.g. in resilience to content changes and in success rate on the largest closed world. The success rates of WF attacks proved to depend on the closed world size, the amount of training data available to the adversary and the computational resources that can be used to train the classifier. For the evaluations performed in this paper, we used the resources available at our institution, but we acknowledge that a more powerful attacker could most likely further improve the attack by using more resources for data collection, model selection and training.

## VI. DISCUSSION

In this section, we enumerate the limitations of this work and discuss remaining open challenges with regard to both the threat model and the deep learning methods we presented.

As in virtually all prior work on WF, we analyzed the attacks only on visits to homepages and omitted other pages within the considered websites. We acknowledge this is an unrealistic assumption. However, as our main goal was to perform a fair comparison with existing attacks, we used the same experimental settings. As the models developed in prior work were tailored to these particular settings, the evaluation of techniques that consider inner web pages was deemed out of scope for this paper. Nevertheless, we find automatic feature learning a promising approach to this problem.

We do not try to approximate the probability of visiting a closed world site vs. a site from the open world in our experiments. We assume that all open world sites have the same prior probability and all closed world sites have the same prior probability. We acknowledge this does not reflect reality but one can only hypothesize on the actual popularity distribution of websites over Tor without risking the privacy of Tor users. It is a limitation of our study and previous work.

Deep learning allows us to replace manual feature engineering with automatic feature learning. Therefore, the resulting attack is not defined by an explicit set of features that would be easily interpretable by a human analyst, but is instead based on abstract implicit non-interpretable features, being learnable parameters of the neural network. Moreover, these features have proven to be more robust to web content changes in comparison to those suggested in prior literature. Consequentially, the corresponding countermeasure cannot focus on concealing specific features as it was done earlier, but in order to defend against the DL-based attack we have to challenge the DL algorithm itself. Therefore, future work should focus on defending against the automated WF attacks, such as deep neural networks presented in this study.

One line of research for future work could be to investigate whether it is possible to mislead the deep neural network predictions. For instance, such research could base on the latest work on *adversarial examples* [6]. These are inputs to the learning model specifically crafted to fool the neural network into classifying them into a wrong class. Adversarial examples can be explored as a defense strategy against DL-based WF in order to protect Tor user's privacy.

In the very recent work by Wang and Goldberg [33], a defense technique based on half-duplex communication and burst molding is proposed. The authors claim that this defense defeats all WF attack techniques known to date. It would be interesting to validate whether the author's claims still hold in the presence of automatic feature learners such as DL.

## VII. CONCLUSION

In this study, we propose a new website fingerprinting attack based on deep learning. The main objective was to assess the feasibility of WF through automated feature learning. We show that deep neural networks are capable of fingerprinting websites with an accuracy that is comparable to the best-performing approaches among numerous research efforts in recent years. The three DNNs we investigated have shown their strengths and weaknesses in the context of WF:

- SDAE performed well overall and proved to be the most stable DNN with respect to the closed world setting.
- CNN is the fastest network due to fewer learnable parameters, and performed best for smaller closed worlds and for the open world evaluation. However, this DNN has a higher risk of overfitting, which was revealed by the larger closed worlds and the concept drift experiments.
- LSTM performed the slowest, but exhibited the best generalization capabilities due to its recurrent structure. However, its constraint in backpropagation did not allow to process long traffic traces without jeopardizing the overall performance.

In certain experimental settings, our attack even improves existing implementations:

- SDAE showed better results than CUMUL on the largest closed world we evaluated.
- All three DL approaches prove to be more robust against web content changes than CUMUL, with LSTM being twice more robust.
- SDAE and CNN networks perform slightly better in the open world evaluation than CUMUL.
- The DL approach is generally more scalable due to parallelization and automated model selection.

In conclusion, using DL gives an adversary major advantages, resulting in accurate and efficient traffic deanonymization.

ACKNOWLEDGMENT

This research is partially funded by the Research Fund KU Leuven. Marc Juarez is funded by a PhD fellowship of the Fund for Scientific Research - Flanders (FWO). We gratefully acknowledge the support of NVIDIA Corporation with the donation of the Titan Xp GPU used for this research.

REFERENCES

[1] K. Abe and S. Goto, "Fingerprinting attack on tor anonymity using deep learning," *Proceedings of the Asia-Pacific Advanced Network*, vol. 42, pp. 15–20, 2016.

[2] J. Bergstra, D. Yamins, and D. Cox, "Making a science of model search: Hyperparameter optimization in hundreds of dimensions for vision architectures," in *Proceedings of the 30th International Conference on Machine Learning*, ser. Proceedings of Machine Learning Research, S. Dasgupta and D. McAllester, Eds., vol. 28, no. 1. Atlanta, Georgia, USA: PMLR, 17–19 Jun 2013, pp. 115–123. [Online]. Available: http://proceedings.mlr.press/v28/bergstra13.html

[3] J. Buchner, "ImageHash," https://github.com/JohannesBuchner/imagehash, 2017.

[4] X. Cai, R. Nithyanand, T. Wang, R. Johnson, and I. Goldberg, "A Systematic Approach to Developing and Evaluating Website Fingerprinting Defenses," in *ACM Conference on Computer and Communications Security (CCS)*. ACM, 2014, pp. 227–238.

[5] X. Cai, X. C. Zhang, B. Joshi, and R. Johnson, "Touching from a Distance: Website Fingerprinting Attacks and Defenses," in *ACM Conference on Computer and Communications Security (CCS)*. ACM, 2012, pp. 605–616.

[6] N. Carlini and D. Wagner, "Towards evaluating the robustness of neural networks," in *IEEE Symposium on Security and Privacy (S&P)*, 2017, pp. 39–57.

[7] M. S. Charikar, "Similarity estimation techniques from rounding algorithms," in *Proceedings of the thiry-fourth annual ACM symposium on Theory of computing*. ACM, 2002, pp. 380–388.

[8] H. Cheng and R. Avnur, "Traffic Analysis of SSL Encrypted Web Browsing," *Project paper, University of Berkeley*, 1998, Available at http://www.cs.berkeley.edu/~daw/teaching/cs261-f98/projects/final-reports/ronathan-heyning.ps.

[9] G. Cherubin, J. Hayes, and M. Juarez, ""Website Fingerprinting Defenses at the Application Layer"," in *Privacy Enhancing Technologies Symposium (PETS)*. De Gruyter, 2017, pp. 168–185.

[10] F. Chollet *et al.*, "Keras," https://github.com/fchollet/keras, 2015.

[11] R. Dingledine, N. Mathewson, and P. F. Syverson, ""Tor: The Second-Generation Onion Router"," in *USENIX Security Symposium*. USENIX Association, 2004, pp. 303–320.

[12] K. P. Dyer, S. E. Coull, T. Ristenpart, and T. Shrimpton, "Peek-a-Boo, I Still See You: Why Efficient Traffic Analysis Countermeasures Fail," in *IEEE Symposium on Security and Privacy (S&P)*. IEEE, 2012, pp. 332–346.

[13] J. Hayes and G. Danezis, "k-fingerprinting: a Robust Scalable Website Fingerprinting Technique," in *USENIX Security Symposium*. USENIX Association, 2016, pp. 1–17.

[14] D. Herrmann, R. Wendolsky, and H. Federrath, "Website Fingerprinting: Attacking Popular Privacy Enhancing Technologies with the Multinomial Naïve-Bayes Classifier," in *ACM Workshop on Cloud Computing Security*. ACM, 2009, pp. 31–42.

[15] G. E. Hinton, S. Osindero, and Y.-W. Teh, "A fast learning algorithm for deep belief nets," *Neural computation*, vol. 18, no. 7, pp. 1527–1554, 2006.

[16] A. Hintz, "Fingerprinting Websites Using Traffic Analysis," in *Privacy Enhancing Technologies (PETs)*. Springer, 2003, pp. 171–178.

[17] S. Hochreiter and J. Schmidhuber, "Long short-term memory," *Neural computation*, vol. 9, no. 8, pp. 1735–1780, 1997.

[18] ——, "Long short-term memory," *Neural computation*, vol. 9, no. 8, pp. 1735–1780, 1997.

[19] M. Juarez, S. Afroz, G. Acar, C. Diaz, and R. Greenstadt, "A critical evaluation of website fingerprinting attacks," in *ACM Conference on Computer and Communications Security (CCS)*. ACM, 2014, pp. 263–274.

[20] M. Juarez, M. Imani, M. Perry, C. Diaz, and M. Wright, "Toward an Efficient Website Fingerprinting Defense," in *European Symposium on Research in Computer Security (ESORICS)*. Springer, 2016, pp. 27–46.

[21] Y. LeCun and Y. Bengio, "The handbook of brain theory and neural networks," M. A. Arbib, Ed. Cambridge, MA, USA: MIT Press, 1998, ch. Convolutional Networks for Images, Speech, and Time Series, pp. 255–258. [Online]. Available: http://dl.acm.org/citation.cfm?id=303568.303704

[22] M. Liberatore and B. N. Levine, ""Inferring the source of encrypted HTTP connections"," in *ACM Conference on Computer and Communications Security (CCS)*. ACM, 2006, pp. 255–263.

[23] V. Nair and G. E. Hinton, "Rectified linear units improve restricted boltzmann machines," in *Proceedings of the 27th International Conference on Machine Learning (ICML-10)*, J. Frnkranz and T. Joachims, Eds. Omnipress, 2010, pp. 807–814. [Online]. Available: http://www.icml2010.org/papers/432.pdf

[24] A. Panchenko, F. Lanze, A. Zinnen, M. Henze, J. Pennekamp, K. Wehrle, and T. Engel, "Website fingerprinting at internet scale," in *Network & Distributed System Security Symposium (NDSS)*. IEEE Computer Society, 2016, pp. 1–15.

[25] A. Panchenko, L. Niessen, A. Zinnen, and T. Engel, "Website fingerprinting in onion routing based anonymization networks," in *ACM Workshop on Privacy in the Electronic Society (WPES)*. ACM, 2011, pp. 103–114.

[26] N. Srivastava, G. E. Hinton, A. Krizhevsky, I. Sutskever, and R. Salakhutdinov, "Dropout: a simple way to prevent neural networks from overfitting." *Journal of Machine Learning Research*, vol. 15, no. 1, pp. 1929–1958, 2014.

[27] Q. Sun, D. R. Simon, and Y. M. Wang, "Statistical Identification of Encrypted Web Browsing Traffic," in *IEEE Symposium on Security and Privacy (S&P)*. IEEE, 2002, pp. 19–30.

[28] Theano Development Team, "Theano: A Python framework for fast computation of mathematical expressions," *arXiv e-prints*, vol. abs/1605.02688, May 2016. [Online]. Available: http://arxiv.org/abs/1605.02688

[29] P. Vincent, H. Larochelle, I. Lajoie, Y. Bengio, and P.-A. Manzagol, "Stacked denoising autoencoders: Learning useful representations in a deep network with a local denoising criterion," *Journal of Machine Learning Research*, vol. 11, no. Dec, pp. 3371–3408, 2010.

[30] T. Wang, X. Cai, R. Nithyanand, R. Johnson, and I. Goldberg, "Effective Attacks and Provable Defenses for Website Fingerprinting," in *USENIX Security Symposium*. USENIX Association, 2014, pp. 143–157.

[31] T. Wang and I. Goldberg, "Improved Website Fingerprinting on Tor," in *ACM Workshop on Privacy in the Electronic Society (WPES)*. ACM, 2013, pp. 201–212.

[32] ——, "On realistically attacking tor with website fingerprinting," in *Proceedings on Privacy Enhancing Technologies (PoPETs)*. De Gruyter Open, 2016, pp. 21–36.

[33] ——, "Walkie-talkie: An efficient defense against passive website fingerprinting attacks," in *26th USENIX Security Symposium (USENIX Security 17)*. Vancouver, BC: USENIX Association, 2017, pp. 1375–1390. [Online]. Available: https://www.usenix.org/conference/usenixsecurity17/technical-sessions/presentation/wang-tao

[34] Z. Wang, "The applications of deep learning on traffic identification," *BlackHat USA*, 2015.

## Appendix

This section elaborates further on the DNN models and learning algorithms we used in our WF attack.

### A. Stacked Denoising Autoencoder

*Autoencoder* (AE) is a shallow feedforward neural network designed for learning meaningful data representations [29]. It is composed of an input layer, one hidden layer and an output layer, as shown in Figure 8a. The input layer acts as an encoder that transforms data and passes it to the hidden layer $\mathbf{h} = f(\mathbf{x})$, and the output layer of the same size acts as a decoder that reconstructs the data back from the hidden layer $\mathbf{r} = g(\mathbf{h})$, intending to produce maximally similar values.

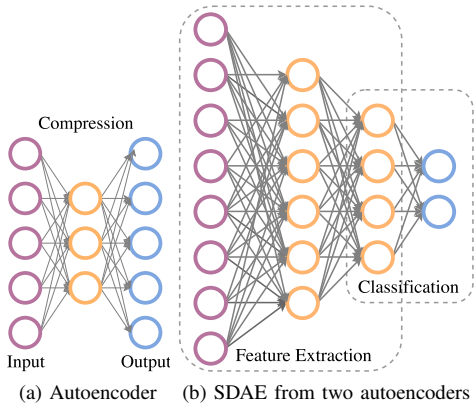

Fig. 8: Stacked Denoising Autoencoder

(a) Autoencoder  (b) SDAE from two autoencoders

The size of the hidden layer plays a crucial role in the AE's working algorithm: it defines the representation of the input used for reconstructing the data. The hidden layer $\mathbf{h}$ is constrained to have fewer neurons than the input $\mathbf{x}$. Then such an *undercomplete* AE is forced to compress the input and can only output its approximation rather than the identity. In order to reconstruct the data from a compressed representation with a minimal loss, the network has to prioritize between properties of the data during compression.

In case of a traffic trace as an input, AE will learn certain combinations and transformations of the input values that allow to reconstruct the same trace with the highest accuracy. As a result, the hidden layer will contain the most salient features of the traffic trace. The training is performed by backpropagating the reconstruction errors expressed via the loss function that has to be optimized by the network. The loss function $L(\mathbf{x}, g(f(\mathbf{x})))$, such as mean squared error, reflects the difference between the input $\mathbf{x}$ and its reconstruction $g(f(\mathbf{x}))$, and reaches its minimum value in case of a total similarity between the two. We use a mean squared error for this purpose, which measures the average of the squares of the deviations: $L(\mathbf{x}, g(f(\mathbf{x}))) = \frac{1}{N} \sum_{i=1}^{N} (g(f(x_i)) - x_i)^2$, where $N$ is the number of neurons of the input (and the output) layer.

Since the undercomplete AE cannot learn a total identity function but only an approximation, its training stops once having minimized the loss function, and thus ensures a good learned representation of data. The AE, as a building block of our future classifier, has to learn representations which reflect statistical properties of the whole data distribution beyond the training examples. This is necessary to achieve a high performance of the model on unseen data, a property of the machine learning models known as a *generalization* capability. The AE that performs during training much better than on traffic unseen before, has *overfitted* to the training data, and thus shows poor generalization capabilities.

To ensure generalization, we apply *regularization* by using *dropout*, when a randomly chosen fraction of input values is set to 0 at each training iteration. AE with dropout is a *Denoising Autoencoder* (DAE) which is more robust to overfitting [26].

*Stacked Denoising Autoencoder* is a deep feedforward neural network built from multiple DAEs by stacking them together, in a manner depicted in Figure 8b. SDAE stacks the DAEs representation layers: the hidden layer of the first DAE is used as the input layer of the successive DAE, and so forth. Chaining several DAEs enables the model to hierarchically extract data from the input to learn features of different levels of abstraction. We chain 3 DAEs to form a 5-layered SDAE. Deeper models produce final features of higher abstraction, which are meant to be used for classification on the concluding layer. The classification layer has one neuron for each possible class, or in our case for each website. Output neurons compute the probability of the input instance to belong to a class. The neuron that produced a maximum probability assigns its label to the training instance.

It was discovered by Hinton et al.[15] that in order to achieve a better performing DNN, it has to first be pre-trained in an unsupervised fashion, that is without using the knowledge of labels of the training data. This strategy is known as the *greedy layer-wise unsupervised pretraining* that initializes the SDAE. This stage is followed by a *supervised fine-tuning* of the whole model, that learns to classify the input by backpropagating the classification errors. The loss function that expresses the errors is a categorical entropy $E = -\frac{1}{N} \sum_{i}^{N} (p_i log_2 p_i)$, where $p_i$ is a returned probability for the predicted class with $N$ websites in total. A classifier confident of its decisions gives a high probability for each predicted class which results into a minimized entropy.

### B. Convolutional Neural Network

A deep network called *Convolutional Neural Network* (CNN) is another feedforward network trained with backpropagation similarly to SDAE, but has a different structure, designed for minimal preprocessing [21]. CNN's main building block is a convolutional layer, which performs a linear *convolution* operation instead of a regular matrix multiplication. The learnable parameters of the convolutional layers are *kernels* or *filters* – multidimensional arrays that are convolved with the input data to create *feature maps*, as depicted in Figure 9. The kernel is applied spatially to small regions of the input, thus enabling sparse connectivity and reducing the actual parameter learning in comparison to fully-connected layers. The kernel aims to learn an individual part of an underlying feature set, e.g. the website fingerprint in a traffic trace. The convolution function is followed by a non-linear activation, typically a *rectifier* [23]. The rectified feature maps are stacked together along the depth dimension to produce the output.

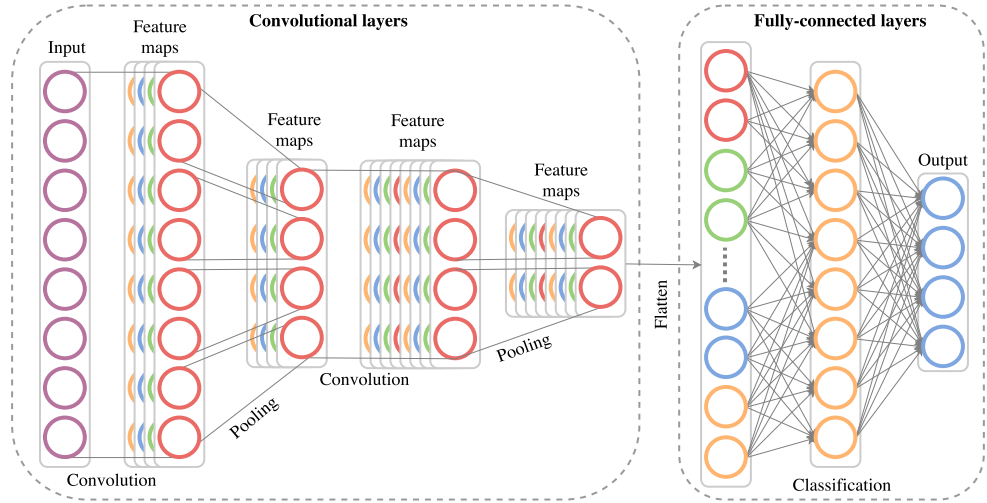

Fig. 9: Convolutional Neural Network.

The next operation of the CNN is typically a *pooling* layer that performs a subsampling operation by replacing the output of the convolution layer with a summary statistics of the nearby outputs. We use a max pooling layer that reports the maximum outputs within regions of the feature maps. Pooling helps the representation become invariant to minor changes of the input. For instance, such subsampling allows to find the prominent identifying parts of the website fingerprint within the traffic trace, despite its slight shifts in location and ignoring the surrounding traffic.

The network can include a whole series of convolution and pooling layers in order to extract more abstract features. We use two sets of such layers. The resulting feature maps need to be flattened and concluded by at least one regular fully-connected layer prior to classification. Because of the risk of overfitting, we apply dropout and limit the amount of learnable parameters of the network by using only two fully-connected hidden layers. The final layer outputs the predictions.

*C. Long Short Term Memory*

*Recurrent neural network* (RNN) is a network with feedback connections, which enable it to learn temporal dependencies [17]. RNN can interpret the input as a sequence, taking into account its temporal properties.

*Long short term memory network* (LSTM) [18] shown in Figure 10a is a special type of a RNN that accommodates so-called LSTM building block to model long-term memory, which allows the network to learn longer input sequences.

The LSTM block processes sequences time step by time step, passing the data through its *memory cells*, and input, output and forget *gates*, as depicted in Figure 10b.

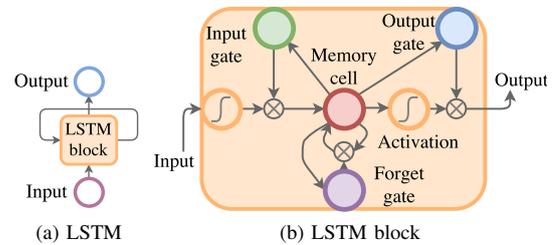

(a) LSTM     (b) LSTM block

Fig. 10: Long Short Term Memory

The memory cell represents the so-called *internal state* of the network. LSTM is able to remove or add information to the cell, regulating these operations by gates. Gates are composed of a sigmoid neural network layer and a pointwise product, and are parameterized by a set of learnable weights. Gates learn to carefully choose whether to let the information through them in order to modify the internal state, to forget information or to produce the output when deemed necessary. The output of an LSTM block is formed by the number of memory units.

LSTM layer's depth depends on the length of processed sequences: due to the feedback connection, they basically have one layer for every processed time step of a sequence. Such structure can be obtained by unrolling the loop in Figure 10a. Classification errors are backpropagated through many layers "through time", which limits the training process: first it significantly slows down training in compare to the feedforward networks, and secondly, in practice it only allows to backpropagate up to 100-200 layers.

LSTM layers can be stacked to form deeper networks. The intuition is the same that higher LSTM layers can capture more abstract concepts. We chain two hidden LSTM layers and form a 4-layered LSTM network (with each layer "unrolled" to as many layers as there are time steps in the processed sequence), which allowed to obtain the best performance.